\documentclass{pasj01}

\RequirePackage{mathpazo}
\RequirePackage[T1]{fontenc}

\newcounter{author}
\def\authorcount#1#2{\refstepcounter{author}\label{#1}
                     \altaffiltext{\ref{#1}}{#2}}

\begin{document}
\SetRunningHead{T. Kato et al.}{Standstills in SU UMa-Type Dwarf Nova NY Ser}

\Received{201X/XX/XX}
\Accepted{201X/XX/XX}

\title{Discovery of Standstills in the SU UMa-Type Dwarf Nova
       NY Serpentis}

\author{Taichi~\textsc{Kato}\altaffilmark{\ref{affil:Kyoto}*}
        Elena~P.~\textsc{Pavlenko},\altaffilmark{\ref{affil:CrAO}}
        Nikolaj~V.~\textsc{Pit},\altaffilmark{\ref{affil:CrAO}}
        Kirill~A.~\textsc{Antonyuk},\altaffilmark{\ref{affil:CrAO}}
        Oksana~I.~\textsc{Antonyuk},\altaffilmark{\ref{affil:CrAO}}
        Julia~V.~\textsc{Babina},\altaffilmark{\ref{affil:CrAO}}
        Aleksei~V.~\textsc{Baklanov},\altaffilmark{\ref{affil:CrAO}}
        Aleksei~A.~\textsc{Sosnovskij},\altaffilmark{\ref{affil:CrAO}}
        Sergey~P.~\textsc{Belan},\altaffilmark{\ref{affil:CrAO}}
        Yutaka~\textsc{Maeda},\altaffilmark{\ref{affil:Mdy}}
        Yuki~\textsc{Sugiura},\altaffilmark{\ref{affil:OKU}}
        Sho~\textsc{Sumiya},\altaffilmark{\ref{affil:OKU}}
        Hanami~\textsc{Matsumoto},\altaffilmark{\ref{affil:OKU}}
        Daiki~\textsc{Ito},\altaffilmark{\ref{affil:OKU}}
        Kengo~\textsc{Nikai},\altaffilmark{\ref{affil:OKU}}
        Naoto~\textsc{Kojiguchi},\altaffilmark{\ref{affil:Kyoto}}$^,$\altaffilmark{\ref{affil:OKU}}
        Katsura~\textsc{Matsumoto},\altaffilmark{\ref{affil:OKU}}
        Pavol~A.~\textsc{Dubovsky},\altaffilmark{\ref{affil:Dubovsky}}
        Igor~\textsc{Kudzej},\altaffilmark{\ref{affil:Dubovsky}}
        Tom\'a\v{s}~\textsc{Medulka},\altaffilmark{\ref{affil:Dubovsky}}
        Yasuyuki~\textsc{Wakamatsu},\altaffilmark{\ref{affil:Kyoto}}
        Ryuhei~\textsc{Ohnishi},\altaffilmark{\ref{affil:Kyoto}}
        Takaaki~\textsc{Seki},\altaffilmark{\ref{affil:Kyoto}}
        Keisuke~\textsc{Isogai},\altaffilmark{\ref{affil:Kyoto}}
        Andrii~O.~\textsc{Simon},\altaffilmark{\ref{affil:TarasShevshenko}}
        Yaroslav~O.~\textsc{Romanjuk},\altaffilmark{\ref{affil:MainUkraine}}
        Alexsandr~R.~\textsc{Baransky},\altaffilmark{\ref{affil:TarasShevshenkoObs}}
        Aleksandr~V.~\textsc{Sergeev},\altaffilmark{\ref{affil:ICAMER}}$^,$\altaffilmark{\ref{affil:Terskol}}
        Vira~G.~\textsc{Godunova},\altaffilmark{\ref{affil:ICAMER}}
        Inna~O.~\textsc{Izviekova},\altaffilmark{\ref{affil:TarasShevshenko}}$^,$\altaffilmark{\ref{affil:ICAMER}}
        Volodymyr~A.~\textsc{Kozlov},\altaffilmark{\ref{affil:ICAMER}}
        Aleksandr~S.~\textsc{Sklyanov},\altaffilmark{\ref{affil:Kazan}}
        Roman~Ya.~\textsc{Zhuchkov},\altaffilmark{\ref{affil:Kazan}}
        Alexei~G.~\textsc{Gutaev},\altaffilmark{\ref{affil:Kazan}}
        Vasyl~O.~\textsc{Ponomarenko},\altaffilmark{\ref{affil:TarasShevshenkoObs}}
        Volodymyr~V.~\textsc{Vasylenko},\altaffilmark{\ref{affil:TarasShevshenko}}
        Ian~\textsc{Miller},\altaffilmark{\ref{affil:Miller}}
        Kiyoshi~\textsc{Kasai},\altaffilmark{\ref{affil:Kai}}
        Shawn~\textsc{Dvorak},\altaffilmark{\ref{affil:Dvorak}}
        Kenneth~\textsc{Menzies},\altaffilmark{\ref{affil:Menzies}}
        Enrique~\textsc{de Miguel},\altaffilmark{\ref{affil:Miguel}}$^,$\altaffilmark{\ref{affil:Miguel2}}
        Stephen~M.~\textsc{Brincat},\altaffilmark{\ref{affil:Brincat}}
        Roger~D.~\textsc{Pickard},\altaffilmark{\ref{affil:BAAVSS}}$^,$\altaffilmark{\ref{affil:Pickard}}
}

\authorcount{affil:Kyoto}{
     Department of Astronomy, Kyoto University, Kyoto 606-8502, Japan}
\email{$^*$tkato@kusastro.kyoto-u.ac.jp}

\authorcount{affil:CrAO}{
     Federal State Budget Scientific Institution ``Crimean Astrophysical
     Observatory of RAS'', Nauchny, 298409, Republic of Crimea}

\authorcount{affil:Mdy}{
     Kaminishiyamamachi 12-14, Nagasaki, Nagasaki 850-0006, Japan}

\authorcount{affil:OKU}{
     Osaka Kyoiku University, 4-698-1 Asahigaoka, Osaka 582-8582, Japan}

\authorcount{affil:Dubovsky}{
     Vihorlat Observatory, Mierova 4, 06601 Humenne, Slovakia}

\authorcount{affil:TarasShevshenko}{
     Astronomy and Space Physics Department, Taras Shevshenko National
     University of Kyiv, Volodymyrska str. 60, Kyiv, 01601, Ukraine}

\authorcount{affil:MainUkraine}{
     Main Astronomical Observatory of the National Academy of Sciences of
     Ukraine, 27 Acad. Zabolotnoho str., Kyiv, 03143, Ukraine}

\authorcount{affil:TarasShevshenkoObs}{
     Astronomical Observatory of Taras Shevshenko
     National University of Kyiv,
     Observatorna str. 3, 04053, Kyiv, Ukraine}

\authorcount{affil:ICAMER}{
     ICAMER Observatory of the National Academy of Sciences of
     Ukraine, 27 Acad. Zabolotnoho str., Kyiv, 03143, Ukraine}

\authorcount{affil:Terskol}{
     Terskol Branch of Institute of Astronomy, Russian Academy of Science, s.
     Terskol, Kabardino-Balkarian Republic, 361605, Russia}

\authorcount{affil:Kazan}{
     Kazan (Volga region) Federal University, Kazan, 420008,
     Kremlyovskaya 18, Russia}

\authorcount{affil:Miller}{
     Furzehill House, Ilston, Swansea, SA2 7LE, UK}

\authorcount{affil:Kai}{
     Baselstrasse 133D, CH-4132 Muttenz, Switzerland}

\authorcount{affil:Dvorak}{
     Rolling Hills Observatory, 1643 Nightfall Drive,
     Clermont, Florida 34711, USA}

\authorcount{affil:Menzies}{
     Center for Backyard Astrophysics (Framingham), 
     318A Potter Road, Framingham, MA 01701, USA}

\authorcount{affil:Miguel}{
     Departamento de Ciencias Integradas, Facultad de Ciencias
     Experimentales, Universidad de Huelva,
     21071 Huelva, Spain}

\authorcount{affil:Miguel2}{
     Center for Backyard Astrophysics, Observatorio del CIECEM,
     Parque Dunar, Matalasca\~nas, 21760 Almonte, Huelva, Spain}

\authorcount{affil:Brincat}{
     Flarestar Observatory, San Gwann SGN 3160, Malta}

\authorcount{affil:BAAVSS}{
     The British Astronomical Association, Variable Star Section (BAA VSS),
     Burlington House, Piccadilly, London, W1J 0DU, UK}

\authorcount{affil:Pickard}{
     3 The Birches, Shobdon, Leominster, Herefordshire, HR6 9NG, UK}



\maketitle

\begin{abstract}
We found that the SU UMa-type dwarf nova NY Ser
in the period gap [orbital period 0.097558(6) d]
showed standstills twice in 2018.
This is the first clear demonstration of a standstill
occurring between superoutbursts of an SU UMa-type
dwarf nova.  There was no sign of superhumps during
the standstill, and at least one superoutburst directly
started from the standstill.  This provides strong evidence
that the 3:1 resonance was excited during standstills.
This phenomenon indicates that the disk radius can
grow during standstills.
We also interpret that the condition close to the limit
of the tidal instability caused early quenching of
superoutbursts, which resulted substantial amount of
matter left in the disk after the superoutburst.
We interpret that the substantial matter in the disk
in condition close to the limit of the tidal instability
is responsible for standstills (as in the high mass-transfer
system NY Ser) or multiple rebrightenings (as in
the low mass-transfer system V1006 Cyg).
\end{abstract}

\section{Introduction}

   Dwarf novae are a class of cataclysmic variables,
which are close binary systems composed of
a white dwarf and a mass-transferring red-dwarf secondary.
The transferred matter forms an accretion disk and
thermal instability in the disk is believed to cause outbursts,
which characterize dwarf novae.
It is known that when the mass-transfer rate ($\dot{M}$)
is above a certain value ($\dot{M}_{\rm crit}$),
the disk is thermally stable and that
the system does not undergo outbursts.  These systems
are called novalike variables.
If the mass-transfer rate is below this, dwarf nova-type
outbursts occur
[for general information of cataclysmic variables
and dwarf novae, see e.g. \citet{war95book}].

   In dwarf novae with small (less than
$\sim$0.25) mass ratios ($q$=$M_2$/$M_1$, where $M_1$
and $M_2$ represent the masses of the white dwarf primary
and the secondary, respectively), the disk radius can
reach the radius of the 3:1 resonance between the binary
rotation and the motion in the disk, and the disk then
becomes tidally unstable and non-axisymmetric deformation
occurs (tidal instability: \cite{whi88tidal};
\cite{hir90SHexcess}; \cite{lub91SHa}), which produces
superhumps [humps with periods a few to several percent
longer than the orbital periods ($P_{\rm orb}$)]
and superoutbursts.  Such systems are called
SU UMa-type dwarf novae and most of then have 
$P_{\rm orb}$ shorter than $\sim$2 hr,
below the period gap in the distribution of $P_{\rm orb}$
of cataclysmic variables.

   Some dwarf novae spend intermediate states between
the outburst maximum and minimum for considerable durations
(tens of days to years), and such states are called
standstills.  Systems showing standstills are called
Z Cam-type dwarf novae.  Z Cam-type dwarf novae usually
have longer orbital periods longer than $\sim$3 hr,
and they are not expected to show SU UMa-type behavior.
It is usually considered that Z Cam-type dwarf novae
have mass-transfer rates close to $\dot{M}_{\rm crit}$
and a subtle variation in the mass-transfer rate
makes the disk in these systems either thermally stable
(standstills) or unstable (dwarf nova state).
Although there is no special reason to believe that
SU UMa-type dwarf novae and Z Cam-type dwarf novae are
completely exclusive, most of known (non-magnetic)
cataclysmic variables can be divided into four regions
on the $\dot{M}$-$q$ plane
based on the thermal and tidal stability criteria,
and Z Cam-type dwarf novae occupy
the region slightly below the thermal stability line
and $q$ higher than the tidal stability line
\citep{osa96review}.

   There has been an exceptional case of BK Lyn, which
had been known as a novalike variable, showed
an SU UMa-type state in 2005--2013
(\cite{pat13bklyn}; \cite{Pdot4}; \cite{Pdot5}).
This object is suspected to be a post-outburst nova
which erupted in AD 101 \citep{pat13bklyn}.
The variation in the activity (mass-transfer rate)
may be affected by the recent nova explosion.

   We found that the SU UMa-type dwarf nova NY Ser
experienced standstills and that the system developed
superoutbursts arising from these standstills
first time in all dwarf novae.
We report on this phenomenon and discuss the implication
on the mechanism of standstills and dwarf nova outbursts.

\begin{figure*}
  \begin{center}
    \FigureFile(130mm,90mm){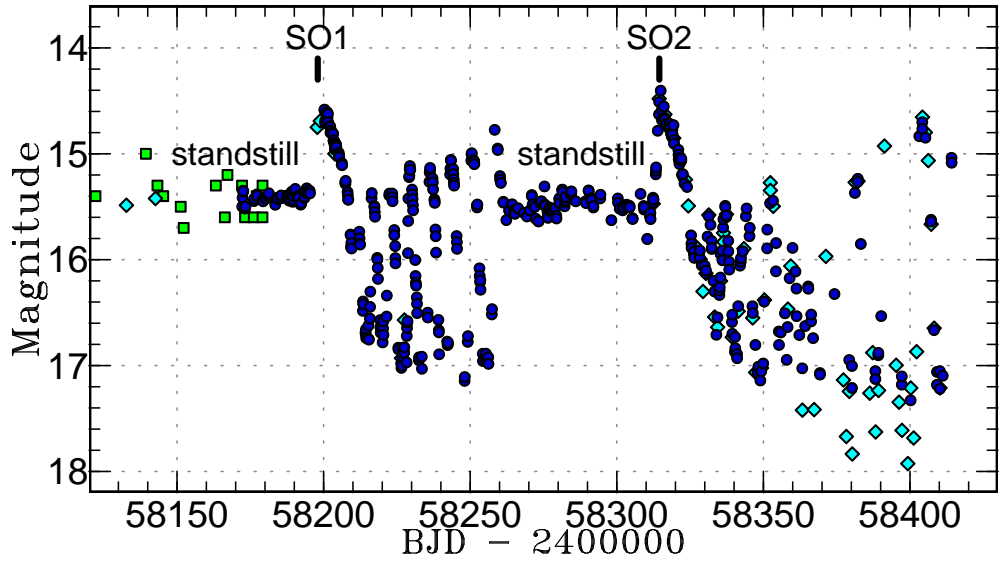}
  \end{center}
  \caption{Light curve of NY Ser in 2018.
  Filled circles, filled square and filled diamonds
  represent our CCD observations, snapshot observations
  by Y.M. and AAVSO $V$-band observations, respectively.
  Two superoutbursts (SO1 and SO2) occurred following
  a standstill.  After superoutbursts,
  the object showed ordinary dwarf nova-type outbursts.
  }
  \label{fig:nyserall2}
\end{figure*}

\begin{figure*}
  \begin{center}
    \FigureFile(130mm,90mm){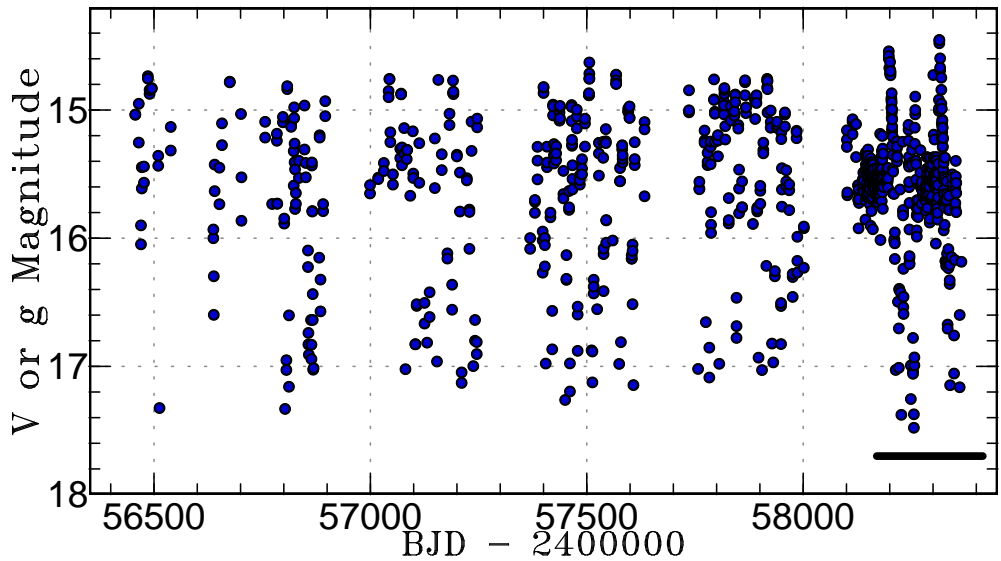}
  \end{center}
  \caption{Long-term light curve of NY Ser between
  2015 and 2018 using All-Sky Automated Survey for Supernovae
  (ASAS-SN) $V$ and $g$ observations
  (\cite{ASASSN}; \cite{koc17ASASSNLC}).
  $V$ and $g$ magnitudes are almost the same in
  dwarf novae and they are not distinguished.
  The final part of this figure (marked by a horizontal bar)
  corresponds to figure \ref{fig:nyserall2}.
  It is very clear that the behavior in 2018 was very different
  from the one in 2015--2017.
  }
  \label{fig:nyserlong}
\end{figure*}

\section{Observations and results}

\subsection{Overall behavior}

   The standstill state was noticed T.K. in 2018 February
in snapshot observations made by Y.M. (vsnet-alert 2189)
and we launched a photometric campaign in
the VSNET Collaboration \citep{VSNET}.
We also used the public data from
the AAVSO International Database\footnote{
   $<$http://www.aavso.org/data-download$>$.
}.
The majority of the data were acquired
by time-resolved CCD photometry by using 20--60cm telescopes
located world-wide.  The data analysis followed
the description in \citet{Pdot} [see E-section 1 in
Supporting information (SI)].  The details of
observations are listed in E-table 1 in SI.
The overall light curve is
shown in figure \ref{fig:nyserall2}.
NY Ser in the past underwent normal outbursts
every 6 to 9~d \citet{nog98nyser} with occasional
superoutbursts.\footnote{
  Although supercycles (intervals between successive superoutbursts)
  were documented to be 85-100~d in \citet{nog98nyser},
  recent observations found that superoutbursts occur more
  irregularly and there are sometimes ``long normal'' outbursts 
  without superhumps \citep{pav14nyser}.
}  We show a long-term light curve in figure \ref{fig:nyserlong}.
It is very clear that the behavior in 2018 was very different
from the one in 2015--2017.

   The standstill state, at least starting on 2018 January 20,
lasted at least until 2018 March 17.
No superhumps were detected during this standstill
(vsnet-alert 21972) and the phenomenon was that of
a genuine standstill of the Z Cam-type dwarf nova,
rather than superhumping novalike state such BK Lyn.

   The object then underwent brightening on
2018 March 22 (it was confirmed to be already in
outburst state on 2018 March 20 based on snapshot
observations from the AAVSO database)
and superhumps appeared
(vsnet-alert 22026).  Following this superoutburst
(rapidly faded on 2018 April 3--4),
the object entered a phase of repetitive
dwarf nova-type outbursts with intervals of 6--7 d
(vsnet-alert 22065),
which lasted at least up to 2018 May 13.
The object was again found to be in a standstill
on 2018 May 24 (vsnet-alert 22195).
Using all the data, the object showed seven
outbursts before entering this standstill.
This standstill lasted up to 2018 July 11
and brightening accompanied by developing superhumps
was recorded on 2018 July 12 (vsnet-alert 22313).
We provided an enlargement of standstill to SO2
in E-figure 1.  There were no gap in observations
longer than 24~hr and a precursor outburst
as in the Kepler observations of V1504 Cyg
(\cite{osa13v1504cygKepler}, in particular figure 4)
and V344 Lyr \citep{osa13v344lyrv1504cyg} can be
safely excluded.
This observation made the first clear evidence
that a superoutburst can start from a standstill.

\subsection{Superhumps}\label{sec:SH}

   During the superoutburst SO1, superhumps were
unambiguously detected (see E-figure 2 in SI).
Since the details of superhump
timing analysis is not very much related to the main
theme of this Letter, we list the times of maxima
in E-table 2 in SI.
It may be, however, worth noting that long superhump
period of 0.10516(3)~d was detected during
the post-superoutburst phase (with dwarf nova outbursts)
between BJD 2458213 and 2458222 (see E-figure 3 in SI).
The fractional superhump excess (in frequency)
$\epsilon^* \equiv 1-P_{\rm orb}/P_{\rm SH}$,
where $P_{\rm SH}$ is the superhump period, was very large
[0.0723(2)] using $P_{\rm orb}$=0.097558(6) d
in \citet{skl18nyser}.  If this long superhump period
reflects the dynamical precession rate
(i.e. pressure effect is negligible,
e.g. \cite{kat13qfromstageA}; E-section 2 in SI),
the $q$ value is estimated to be 0.23 (assuming the maximum
disk size at the 3:1 resonance 0.448$A$,
where $A$ is the binary separation) and 0.43 for (0.35$A$,
typical SU UMa-type dwarf novae in the post-superoutburst
phase; this condition corresponds
to subsection 4.3.2 in \cite{kat13qfromstageA}).
The true $q$ would be something between these two
extreme values.  Since $q$ is expected to lower than
$\sim$0.25 to enable superhumps, a post-superoutburst
disk larger than typical SU UMa-type dwarf novae
is preferred.

   Superhumps during the superoutburst SO2 were less
well observed, although the developing phase of
the superoutburst was observed (E-figure 4 in SI).
It is difficult to tell the superhump period with
nightly coverage of only one orbital cycle
(due to the small solar elongation).
The data were, however, sufficient to tell that
the humps during this superoutburst cannot be expressed
by the orbital period (i.e. these humps were superhumps,
and not enhanced orbital humps).

\section{Discussion}

\subsection{Standstills in SU UMa-type dwarf novae}

   Our observations demonstrated that an SU UMa-type
dwarf nova can have a standstill during a supercycle,
and that superoutbursts can directly occur from standstills.
Such a phenomenon has never been recorded.

   The most important finding is that the absence of
superhumps or negative superhumps during standstills
(see E-figure 5 in SI).
These standstills are indeed of purely Z Cam-type ones
rather than ``permanent superhump'' states seen in
novalike variables such as BK Lyn.
This indicates that the standstills in NY Ser were
not maintained by extra tidal torques produced by
tidal instability nor a result of a disk tilt
(which is supposed to be the cause of negative
superhumps).

   In Z Cam stars, long-term variation in the mass-transfer
rate is considered to be the cause of standstills
\citep{mey83zcam}.  We tested this possibility by
comparing the mean brightness (averaged in the flux unit,
all having standard errors of less than 0.01 mag)
between the outbursting and standstill states.
The mean values were 15.41 (before BJD 2458196,
standstill), 15.46 (BJD 2458196--2458261, SO1 and
the following outburst state), 15.52 (BJD 2458261--2458312,
standstill) and 15.57 (after BJD 2458312, SO2 and
the following outburst state).  There was no systematic
brightness increase during standstills comparable to
0.2 mag difference in some Z Cam stars in \citet{hon98zcam},
and we can exclude the possibility of the increased
mass-transfer as the cause of standstills.
The mean brightness, however, gradually faded,
probably reflecting the long-term
decrease of the mass-transfer rate returning from
the unusual state in 2018 back to normal.

   Superhumps were newly excited at the onset of
each superoutburst.  Following the standard
thermal-tidal instability model
(\cite{osa89suuma}; \cite{osa96review})
this finding indicates that the disk radius gradually
increased during standstills and then eventually reached
the 3:1 resonance.  This observation provides the first
evidence that the disk radius can increase during
standstills.

   Although we are not aware of the precise mechanism
of the increase of the disk radius, we consider that
the re-distribution of the disk matter after
the superoutburst may be responsible for the phenomenon.
We should note that a similar transition from
dwarf nova-type outbursts to a (quasi-)standstill was
recorded during the post-superoutburst state in
the WZ Sge-type dwarf nova
ASASSN-15po \citep{nam17asassn15po}.  A similar case
was found in the helium dwarf nova V803 Cen in 2016
(K. Isogai et al. in preparation).

   We recently found that outbursts arising from standstills
are more prevalent in Z Cam stars [we called such objects
IW And stars, currently a subclass of Z Cam stars \citet{kat18iwandtype}].
Following the present evidence in NY Ser that the disk
radius can increase during standstills, we suggested
in \citet{kat18iwandtype}
that the increase in the disk radius or re-distribution of
the disk matter during standstills, might be the cause of
such a phenomenon.
We consider that this phenomenon is more prevalent
in various kinds of dwarf novae and should require
more theoretical studies.

\subsection{Behavior near limit of tidal stability}

   NY Ser is an object in the period gap
(\cite{nog98nyser}; \cite{skl18nyser}).
It has been shown that ``long-normal'' outbursts
(long outbursts without superhumps) were recorded
in this object \citep{pav14nyser}.  The only other known
SU UMa-type dwarf novae showing long-normal outbursts are TU Men,
(\cite{sto84tumen}; \cite{bat00tumen}) and V1006 Cyg
(\cite{kat16v1006cyg}; \cite{pav18v1006cyg}).

   NY Ser, TU Men and V1006 Cyg have long orbital periods
(longer than 0.095~d), and $q$ values are expected to be
close to 0.25, which is the limit to allow
tidal instability to develop.  In such conditions,
tidal instability is expected to be difficult to develop
or maintain in these systems.  This condition would enable
long-normal outbursts to occur (the disk mass is
enough to sustain a superoutburst while tidal instability
fails to develop).  The same condition may enable easy
decoupling between thermal and tidal instabilities
as proposed in very low $q$ systems \citep{hel01eruma}.
We have already suggested that in systems near the border
of tidal instability is not strong enough to maintain 
the disk in the hot state
when the cooling wave starts \citep{kat16v1006cyg}.

   Early quenching of tidal instability and subsequent
quenching of a superoutburst may result large amount
of matter left in the disk after the superoutburst.
This interpretation appears to be strengthened
by the recent discoveries of SU UMa-type dwarf novae
above the period gap (ASASSN-18yi and ASASSN-18aan,
Wakamatsu et al. in preparation), whose multiple
post-superoutburst rebrightenings are a signature
of the large amount of matter after superoutbursts.
The cause of standstill-like behavior in ASASSN-15po
and V803 Cen may be this decoupling between thermal and
tidal instabilities since both objects are expected
to have very low $q$ values.

   In the case of long-period systems near the border
of tidal instability,
the similar decoupling at the high $q$ limit can happen
and may cause substantial disk matter to remain
even after the superoutburst.  If the mass-transfer rate
from the secondary is low, this condition would enable
to mimic the WZ Sge-type phenomenon (multiple rebrightenings)
as in V1006 Cyg \citep{kat16v1006cyg}.
With a high mass-transfer rate close to the thermal
stability, a system like NY Ser can mimic the Z Cam-type
behavior by enabling a quasi-steady state produced
by the large disk matter in a post-superoutburst disk
combined with the high mass-transfer rate.
This interpretation appears to be
consistent with the large post-superoutburst disk
as inferred from the superhump period
in subsection \ref{sec:SH}.

\section*{Acknowledgements}

We acknowledge with thanks the variable star
observations from the AAVSO International Database contributed by
observers worldwide and used in this research.
We also acknowledge Mayaky observational team
(V. V. Troianskyi and V. I. Kashuba) for
the short term observational support.
This work was performed with the use of observational data
obtained at North-Caucasus astronom ical station of KFU.
The work is partially performed according to the Russian
Government Program of Competitive Growth of Kazan Federal
University.
This work was funded by the subsidy allocated to Kazan Federal
University for the state assignment in the sphere of
scientific activities 3.9780.2017/8.9.
The observations made by Vihorlat Observatory team were
supported by the Slovak Research and Development
Agency under the contract No. APVV-15-0458.
We are grateful to the ASAS-SN team for
making their data available to the public.

\section*{Supporting information}

Additional supporting information can be found in the online version
of this article: Tables.
Figures.\\
Supplementary data is available at PASJ Journal online.

\end{document}